\documentclass[lettersize,journal]{IEEEtran}

\usepackage{amsmath,amsfonts,amssymb}
\usepackage{algorithm}
\usepackage{array}
\usepackage[caption=false,font=normalsize,labelfont=sf,textfont=sf]{subfig}
\usepackage{textcomp}
\usepackage{stfloats}
\usepackage{url}
\usepackage{verbatim}
\usepackage{graphicx}
\usepackage{cite}
\usepackage{multirow, bm}
\usepackage{booktabs}
\usepackage{makecell}
\usepackage{color,soul}
\usepackage{hyperref} 
\usepackage{adjustbox}
\usepackage{algpseudocode}

\begin{document}

\title{Multi-channel Conversational Speaker Separation via Neural Diarization}
\author{Hassan Taherian, and DeLiang Wang,~\IEEEmembership{Fellow,~IEEE}
  \thanks{H. Taherian is with the Department of Computer Science and Engineering, The Ohio State University, Columbus, OH 43210-1277 USA (e-mail: taherian.1@osu.edu).}
  \thanks{D. L. Wang is with the Department of Computer Science and Engineering and the Center for Cognitive and Brain Sciences, The Ohio State University, Columbus, OH 43210-1277 USA. (e-mail: dwang@cse.ohio-state.edu).}}

\maketitle

\begin{abstract}

  When dealing with overlapped speech, the performance of automatic speech recognition (ASR) systems substantially degrades as they are designed for single-talker speech. To enhance ASR performance in conversational or meeting environments, continuous speaker separation (CSS) is commonly employed. However, CSS requires a short separation window to avoid many speakers inside the window and sequential grouping of discontinuous speech segments. To address these limitations, we introduce a new  multi-channel framework called ``speaker separation via neural diarization" (SSND) for meeting environments. Our approach utilizes an end-to-end diarization system to identify the speech activity of each individual speaker. By leveraging estimated speaker boundaries, we generate a sequence of  embeddings, which in turn facilitate the assignment of speakers to the outputs of a multi-talker separation model. SSND addresses the permutation ambiguity issue of talker-independent speaker separation during the diarization phase  through location-based training, rather than during the separation process. This unique approach allows multiple  non-overlapped speakers to be assigned to the same output stream, making it possible to efficiently process long segments—a task impossible with CSS. Additionally, SSND is naturally suitable for speaker-attributed ASR. We evaluate our proposed diarization and separation methods on the open LibriCSS dataset, advancing state-of-the-art diarization and ASR results by a large margin.

\end{abstract}


\section{Introduction} \label{section_1}

\IEEEPARstart{T}{alker-independent} speaker separation systems are increasingly tailored to address more realistic scenarios~\cite{wang2018supervised}. One such environment is conversational or meeting settings. Conversational speech is  characterized by its extended duration, an arbitrary number of participating speakers, and varying degrees of speech overlap.
To address the challenges posed by conversational speech, the notion of continuous speaker separation (CSS) has been introduced~\cite{chen2020continuous}. CSS is designed to process long audio recordings and manage overlapped speech involving an arbitrary number of speakers. In the CSS approach, an audio recording is broken into shorter, partially overlapping segments, typically ranging from 2-3 seconds. Each segment should contain at most two speakers. By doing so, the CSS task is simplified to a two-talker concurrent speaker separation task for each segment.
During the separation process, each segment is treated independently, resulting in two estimated speech signals. In cases where segments contain no overlapped speech, the processing reduces to speech enhancement, and the enhanced signal is mapped to one of the two streams, while the other generates a zero signal. As CSS produces two speech estimates for each segment, there is a requirement to group the estimates of the current segment with those in the previous segment. Sequential grouping, also referred to as ``stitching", is essential for handling same-talker speech that spans multiple segments. This is commonly achieved by comparing the separation results in the overlapped regions between consecutive segments.

Since its inception, the CSS framework has been the subject of numerous studies aiming to enhance various components~\cite{taherian2021time, DontShootButterfly, Chenda, wang2021multi, taherian2023multi, taherian23_interspeech}. In~\cite{taherian2021time}, a modulation factor based on the segment overlap ratio is introduced to dynamically adjust a separation loss. Chen et al.\cite{DontShootButterfly} proposed an early exit mechanism in Transformer layers for multi-channel speaker separation, speeding up inference by exiting when successive layer outputs are similar. Li et al.\cite{Chenda} proposed a dual-path separation model that leverages inter-segment information through a memory embedding pool.  Wang et al.\cite{wang2021multi} employed a multi-stage strategy, combining a multi-input single-output (MISO) separation model with deep learning based beamforming followed by a post-filtering network. Extending upon this approach, the study in\cite{taherian23_interspeech} integrates multi-input multi-output (MIMO) separation, incorporating a multi-resolution loss~\cite{taherian2023multi}.

Despite the significant progress, the CSS framework faces several challenges.
The first challenge is its limited segment size, stemming from the requirement that each segment must contain no more than two speakers~\cite{chen2020continuous}. A shorter segment length creates a bigger difficulty to group the separated utterances of the same talker over a period of time. When processing single-talker segments, a separation model occasionally fails to isolate the speaker in one stream with the other stream silent. Consequently, a speaker is erroneously split into two streams, adversely impacting downstream speech applications like automatic speech recognition (ASR) as they process each stream as originating from a distinct speaker~\cite{taherian2023multi}. To capture longer speech utterances, two recent studies introduced new training criteria based on permutation-invariant training (PIT)~\cite{kolbaek2017multitalker, GraphPIT,zhang22y_interspeech}. In~\cite{GraphPIT}, the authors proposed ``Graph-PIT", an extension of PIT that can separate a varying number of speakers from a two-talker separation model. This model employs graph coloring to optimally assign multiple speakers to two streams. Zhang et al.~\cite{zhang22y_interspeech} introduced Group-PIT, which organizes a long reference signal into utterance groups, employing the PIT criterion for these groups instead of individual utterances.

 The second CSS challenge lies in the stitching process. Typically, the overlap between neighboring segments is set to  50\% of the segment length to provide an adequate context for aligning adjacent segments. 
 However, this approach introduces computational inefficiency since each segment is processed twice. Moreover, spectral distance based stitching is prone to errors, resulting in the misalignment of adjacent segments.

 The third challenge arises when dealing with long recordings, e.g. a one-hour meeting involving many participants. While stitching can group a continuous single-talker signal in consecutive frames, it does not address the challenge of grouping discontinuous signals of the same talker, e.g., how to group the utterance of a talker in the first 5 minutes of a one-hour meeting and the utterance of the same talker in the last five minutes of the meeting. The grouping of discontinuous utterances of the same talker is crucial for subsequent tasks such as speaker-attributed ASR.

 In this paper, we present a new framework, termed ``speaker separation via neural diarization" (SSND), for multi-channel conversational speaker separation.  This approach employs a deep neural network (DNN) for speaker diarization to demarcate the speech activities of individual speakers. Leveraging the estimated utterance boundaries from neural diarization, we generate a sequence of speaker embeddings. These embeddings, in turn, facilitate the assignment of speakers to two output streams of the separation model. The SSND approach tackles the permutation ambiguity issue of talker-independent separation during the diarization phase, rather than during separation. This distinction permits non-overlapped speakers to be assigned to the same output stream, enabling the processing long recordings missing from standard CSS. Furthermore, there is no stitching in SSND, and hence duplicate processing of segments is eliminated, resulting in computational efficiency.  Another advantage of SSND lies in the inherent integration of speaker separation and diarization, enabling sequential grouping of the discontinuous utterances of the same talker.

For embedding extraction, we utilize end-to-end neural diarization with an encoder-decoder-based attractor calculation module (EEND-EDA)\cite{eendEDA}, but extend EEND-EDA to multi-channel scenarios with a different training criterion that can handle a larger number of speakers. Specifically, we propose to use location-based training (LBT)\cite{taherian2022LBTJorunal} to resolve permutation ambiguity in speaker diarization. We show that that LBT significantly outperforms the PIT criterion for diarization of many speakers. Our SSND framework achieves state-of-the-art diarization and ASR results, surpassing all existing CSS based methods on the open LibriCSS dataset~\cite{chen2020continuous}. 

The rest of the paper is organized as follows. In Section~\ref{sec_algorithm}, we describe our proposed multi-channel diarization model, and the SSND framework. We present the experimental setup in Section~\ref{sec_exp_setup}, and evaluation and comparison  results in Section~\ref{sec_results}. Concluding remarks are provided in Section~\ref{section_conclude}.

\section{Algorithm Description} \label{sec_algorithm}

\subsection{Multi-channel Diarization} \label{sec_diarization}

\begin{figure}[!t]
  \centering
  \includegraphics[width=0.40\textwidth]{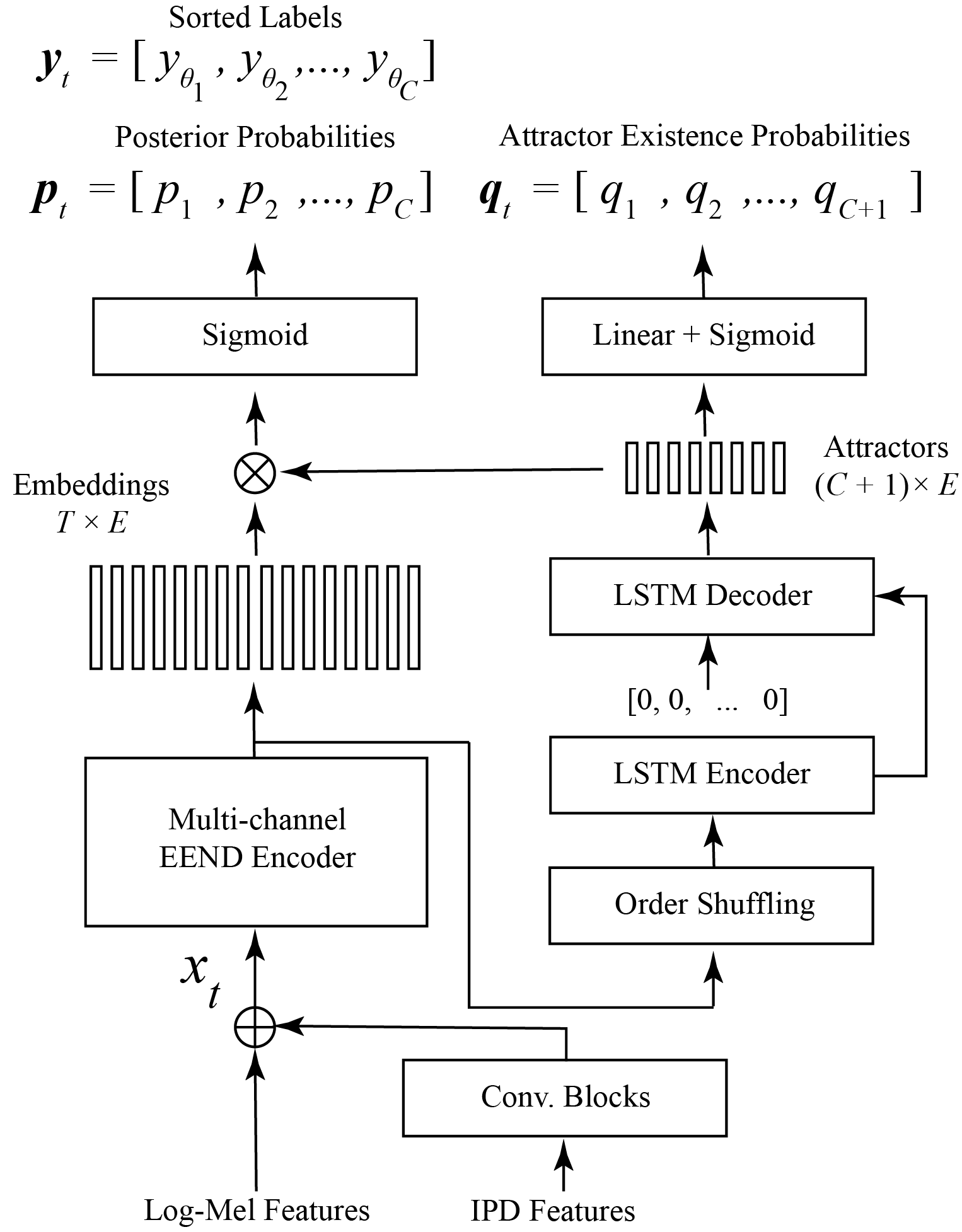}
  \caption{Schematic diagram of the proposed MC-EEND with location-based training. }
  \label{fig_mc_eend}
  \end{figure}

Speaker diarization traditionally employs a clustering-based approach~\cite{diarReview2006,park2022review}. This approach revolves around grouping speaker embeddings, such as x-vectors~\cite{snyder2018x}, into clusters. Such a method usually involves three distinct stages. First, a speech activity detection model is utilized to identify speech intervals. Then, speaker embeddings are extracted, and finally a clustering technique such as spectral clustering (SC) is applied.  However, this method has its limitations. Its stages are independent and cannot be trained jointly to minimize diarization errors. Plus, clustering methods struggle with speaker overlaps, as they assume a single speaker within a segment.

End-to-end neural diarization (EEND) has been introduced to streamline the diarization process by using a single neural network model~\cite{Fujita2019Interspeech}. Unlike clustering-based diarization, the EEND method can handle speaker overlaps. Furthermore, by incorporating an encoder-decoder-based attractor calculation (EDA)~\cite{eendEDA}, EEND can handle an unknown number of speakers. For a sequence of frame-wise  $\boldsymbol{x}_t$, where $t=[1, \dotsc, T]$, EEND encodes these into a series of embeddings, $\boldsymbol{e}_t \in \mathbb{R}^{E} $, via a DNN-based encoder. From these frame-wise embeddings, the EDA module derives a number of attractors $\boldsymbol{a}_c$ for $c=[1, \dotsc, C]$ speakers. Subsequently, speech activity probabilities, $\boldsymbol{p}_t = [{p}_{1}, {p}_{2}, \dotsc, {p}_{C}]$, are derived by taking the dot product of the frame-wise embeddings and the speaker-wise attractors, followed by applying a sigmoidal function. Finally, the speech activities  of different speakers are estimated by using a decision threshold $\tau$.

EEND is formulated for monaural recordings. In this study, we extend EEND-EDA to multi-channel recordings by integrating spatial features. Fig.~\ref{fig_mc_eend} illustrates our proposed multi-channel EEND-EDA (MC-EEND) model. Our model utilizes both spectral and spatial features~\cite{KeSpectrospatial}. For spectral features, we utilize log-Mel filterbanks, while for spatial features, we employ the inter-channel phase difference (IPD) between the reference microphone and the other microphones. For each pair of microphones, the cosine and sine of the IPD are concatenated. These IPD features are then processed through a series of convolutional blocks, each of which is composed of a convolutional layer, a PReLU activation function, and a group normalization layer. Subsequently, the processed IPD features are concatenated with the log-Mel features. The combined features are fed to the EEND encoder, which comprises several Transformer layers without positional encodings. To derive speaker attractors using EDA, we shuffle the order of the embeddings.

For EEND to be effective in real-world applications, it must be trained in a talker-independent manner to accommodate untrained speakers. Much like talker-independent speaker separation, the difficulty lies in aligning diarization output layers with the respective speaker labels. Without proper output-speaker assignment, EEND training would not converge due to conflicting gradients. This is known as the permutation ambiguity problem~\cite{hershey2016deep}, \cite{kolbaek2017multitalker}. Previous studies employ PIT to tackle this problem by analyzing the losses across all possible output-speaker pairings~\cite{Fujita2019Interspeech, eendEDA}. However, unlike speaker separation where the number of concurrent speakers can be reasonably limited to two or three, diarization may involve many speakers. Using PIT in such cases becomes problematic as it has factorial or polynomial training complexity~\cite{tachibana2021towards,dovrat2021many }, making it inefficient for a large number of speakers.

\begin{figure*}[!t]
  \centering
  \includegraphics[width=0.90\textwidth]{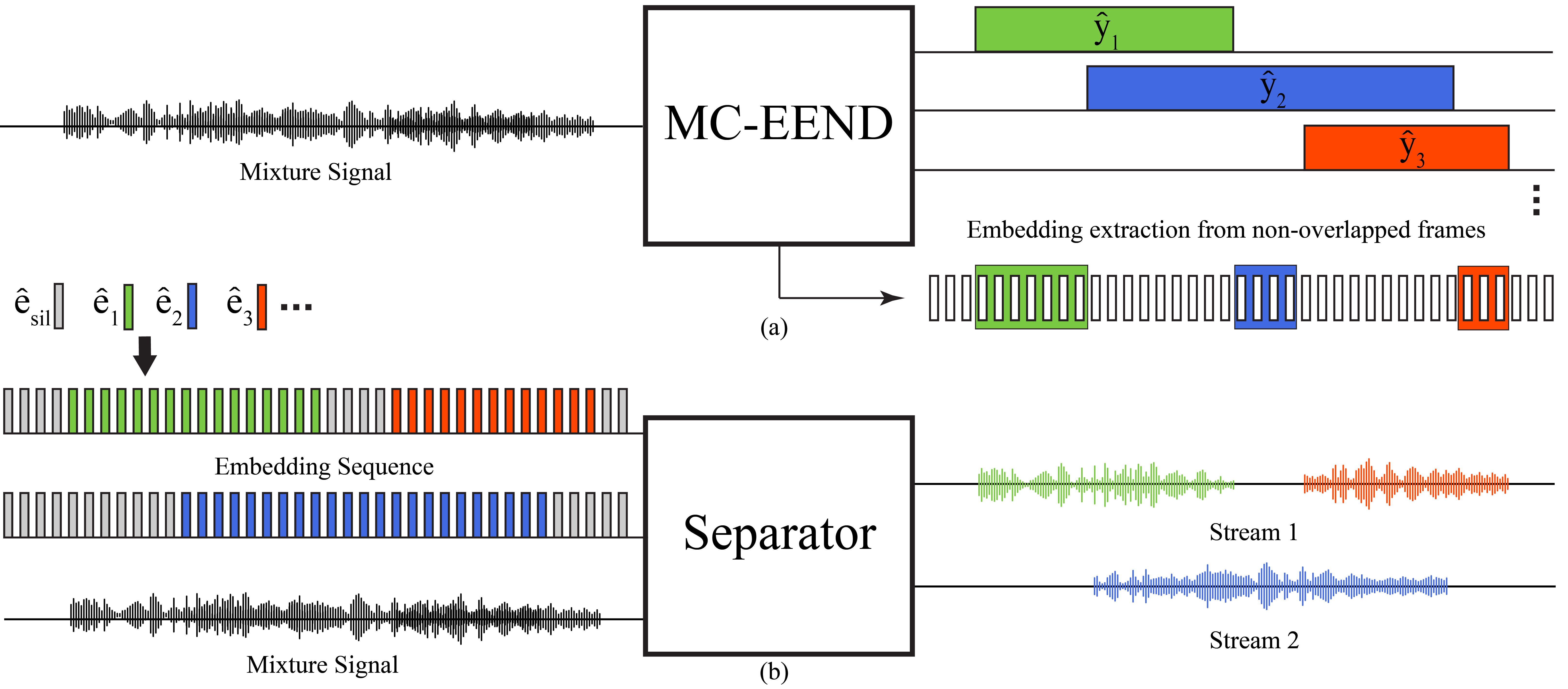}
  \caption{Proposed SSND framework: (a) embedding extraction with MC-EEND based on estimated non-overlapped frames, and (b) constructing embedding sequences and feeding to speaker separation along with mixture signal.  }
  \label{fig_SSND}
  \end{figure*}

In this study, we introduce LBT~\cite{taherian2022LBTJorunal} for diarization. We employ the spatial locations of speakers to determine output-speaker assignments. Given a polar coordinate system with the microphone array's center as the origin, we define the diarization loss function as:

\begin{equation}
  \mathcal{L}_{\text{EEND}} = \frac{1}{TC} \sum_{t=1}^{T} H(\boldsymbol{y}_{t}, \boldsymbol{p}_t)
  \label{eq:diar_loss_LBT}
\end{equation}
where $\boldsymbol{y}_{t} = [{y}_{\theta_1}, {y}_{\theta_2}, \dotsc, {y}_{\theta_C}]$ represents the binary speaker label vector for frame $t$. This vector is arranged in ascending order according to speaker azimuths relative to the microphone array~\cite{taherian2022LBTJorunal}. The binary cross entropy function, $H(.,.)$, is defined as:

\begin{equation}
  H(\boldsymbol{y}_t, \boldsymbol{p}_t) = - \sum_{c=1}^{C} y_{c,t} \log p_{c,t} + (1-y_{c,t}) \log (1-p_{c,t}).
  \label{eq:cross_entropy_func}
  \end{equation}

LBT, or azimuth-based training specifically, has a linear computational complexity~\cite{taherian2022LBTJorunal}, making it efficient to train end-to-end diarization systems with many speakers. 
In the EDA module, attractors are trained using attractor existence probabilities, denoted as $\boldsymbol{q}_t$ (see Fig.~\ref{fig_mc_eend}). The attractor probabilities  are calculated by using a fully connected layer, followed by sigmoidal activation. The EDA module's loss function is described as:

\begin{equation}
  \mathcal{L}_{\text{EDA}} = \frac{1}{C+1} H(\boldsymbol{l}, \boldsymbol{q})
  \label{eq:att_loss}
\end{equation}
where $\boldsymbol{l} \in \mathbb{R}^{C+1} $ is a binary vector with its first $C$ elements set to 1 and its final element set to 0. The total loss for MC-EEDA is expressed as:

\begin{equation}
\mathcal{L_\text{D}} = \mathcal{L}_{\text{EEND}} + \mathcal{L}_{\text{EDA}}.
\label{eq:combined_loss}
\end{equation}

\subsection{Speaker Separation via Neural Diarization (SSND)} \label{sec_SSND}

Our SSND framework uses the MC-EEND diarization model. However, the SSND approach can be coupled with any diarization model. Within this framework, we assume that the number of concurrent speakers at each frame does not exceed two, as more than two speakers rarely talk simultaneously in real-world conversations~\cite{cetin06_interspeech}, \cite{chen2020continuous}. To represent each speaker, we extract speaker embeddings from the MC-EEND encoder. Specifically, for each speaker, we average the embedding vectors over all the frames when the speaker is active with all others silent, which may be discontinuous:

\begin{equation}
  \hat{e}_c = \frac{1}{T_c} \sum_{t} e_{t} \quad \text{where} \quad \left\{ \begin{array}{l}
  \hat{y}_{c,t} = 1, \\
  \hat{y}_{c',t} = 0, ~\text{for}~ c' \neq c.
  \end{array} \right.
  \end{equation}
Here, $\hat{y}_{c}$ represents the estimated speech activities derived from MC-EEND and $T_c$ is the number of frames where only speaker $c$ is active. Fig.~\ref{fig_SSND}(a) illustrates the embedding extraction process.

Based on extracted speaker embeddings, we create two sequences in the following manner. Using the order of diarized speech activities, each speaker-active interval is assigned to one of two embedding sequences. For such an interval, the extracted embedding of the corresponding speaker is assigned to every frame of the interval as illustrated on the left side of Fig.~\ref{fig_SSND}(b). 
We initially assign the first interval to the first embedding sequence.
At the onset of a current interval, if both sequences are silent, we check whether the underlying speaker of the current interval is the same as that of the previously ended interval. If yes, the current interval is assigned to the same sequence as the previously ended sequence; if not, the interval is assigned to the other sequence.  If only one sequence is silent, the current interval is assigned to this sequence. 
This procedure is illustrated in Figure~\ref{fig_SSND}(b). We note that this assignment of speaker-active intervals to two embedding sequences guarantees no overlap between the intervals assigned to each sequence regardless of the number of speakers, as long as no more than 2 talkers speak simultaneously. For silent frames, we use a zero embedding vector $\hat{e}_{sil}$. The resulting two embedding sequences are then fed to a separation network along with the multi-channel mixture signal, as shown in Fig.~\ref{fig_SSND}(b). 

We employ two architectures for the separator network (see Fig.~\ref{fig_SSND}(b)), both of which operate in the short-time Fourier transform (STFT) domain. The first architecture, TF-GridNet~\cite{wang2023tf}, processes time-frequency units in a grid-like manner. TF-GridNet consists of several blocks, each with three main components. The first two components utilize bi-directional long short-term memory (BLSTM) to process full-band spectral features within each frame and the temporal information within each frequency. The last component incorporates a self-attention module, to process full-band information across frames to capture long-range contexts. The second architecture, SpatialNet~\cite{quan2023spatialnet}, processes spectral and temporal information similarly to TF-GridNet but employs only narrow-band and cross-band modules. The narrow-band module employs multi-head self-attention, while the cross-band module uses convolutional layers.

Similar to the embedding sequences, we create two output streams for separated speech signals as illustrated in Fig.~\ref{fig_SSND}(b). Clean speech signals are assigned to a stream based on the corresponding speaker embedding assignment. For  silent frames, we use a zero signal. We train the separator model with an $\ell_1$ norm loss on the real and imaginary components of the estimated signal and target signal at the reference microphone~\cite{williamson2016complex}, with an additional magnitude loss term~\cite{ 9018157}:
\begin{equation}
	\mathcal{L}_{\text{sep}}(\hat{S}, S) = \frac{1}{2} \sum_{n=1}^{2} \mathcal{L}(\hat{S}_n, S_n)
  \label{eq:sep_loss}
\end{equation}

\begin{equation}
  \begin{split}
	\mathcal{L}(\hat{S}, S) &=  \left\lVert \hat{S}^{(r)} - S^{(r)} \right\rVert_{1} + \left\lVert \hat{S}^{(i)} - S^{(i)} \right\rVert_{1} \\
   &+ \left\lVert |\hat{S}| - |S| \right\rVert_{1},
  \end{split}
\end{equation}
where $\hat{S}$ and $S$ denote the estimated and clean speech signals in the STFT domain. Superscripts $r$ and $i$ denote real and imaginary parts, $|~.~|$ computes magnitude, and $ \left\lVert~.~\right\rVert_{1}$ indicates $\ell_1$ norm.

In the proposed SSND framework, the separator network resolves the permutation ambiguity problem by using the embedding sequences for output-speaker assignment. Unlike CSS based on PIT where the number of outputs is equal to the number of speakers, this approach allows multiple non-overlapped speakers to share the same output stream. As a result, SSND can handle long recordings containing numerous speakers. It is worth noting that our approach differs from research on personalized speech enhancement~\cite{OneModel} and speaker separation~\cite{zeghidour2021wavesplit} that use embeddings for speaker extraction. In these studies, speaker embedding does not change over a stream. In contrast, speaker embedding in our approach may change corresponding to a speaker change (e.g., over the first stream of Fig.~\ref{fig_SSND}(b)).

\section{Experimental Setup} \label{sec_exp_setup}

\subsection{Datasets}

We assess the proposed approach for both diarization and multi-speaker ASR tasks using the LibriCSS corpus~\cite{chen2020continuous}. This corpus is structured into ten one-hour sessions. Each session is further subdivided into six ten-minute mini-sessions, which are characterized by varying speech overlap levels. These levels include 0S (no overlap with short pauses ranging from 0.1-0.5 seconds between utterances), 0L (no overlap with long pauses lasting between 2.9-3.0 seconds), and then speaker overlap at 10\%, 20\%, 30\%, and 40\%. Every recording in the dataset was sampled at 16 kHz. The recordings for LibriCSS were drawn from the LibriSpeech development set. To capture real room acoustics, utterances were replayed through loudspeakers, and recorded by a circular microphone array with six microphones positioned in a circle with a radius of 4.25 cm and an additional central microphone. We designate the center microphone as the reference microphone. For speaker diarization and speech recognition, we adopt session-wise evaluation by utilizing every mini-session for assessment, and all 10 sessions are used for evaluation.

To generate training data for diarization, we simulate meeting-style conversations based on a recipe in LibriCSS\footnote{Available online at: \url{https://github.com/jsalt2020-asrdiar/jsalt2020_simulate}}. Each session includes eight speakers, with two utterances for each speaker. These clean speech signals originate from the Librispeech training set~\cite{LibrispeechDataset}. The speaker overlap ratio is chosen randomly from 0 to 45\%. Furthermore, silences varying from 0.5 to 3.0 seconds are inserted between neighboring utterances with a 0.26 probability. On average, the sessions generated for diarization training have a duration of 3 minutes. We employ simulated room impulse responses (RIRs) with reverberation time (T60) chosen in the range of 0.2 and 0.6 seconds. Speaker azimuths are randomly generated and maintain a minimum separation of 5 degrees. We adjust the sound levels between the speakers in the range of -3.5 to 3.5 dB. We also add simulated diffuse noise with the signal-to-noise ratio (SNR) in the range of 10 to 30 dB, where all reverberant speech utterances are considered as the signal in the SNR calculation.

To generate training data for speaker separation, we use the same recipe but opt for shorter sessions. Each session in our data generation has 1-2 utterances per speaker, chosen at random with each utterance under 10 seconds. Overlap ratios are randomly selected between 0.4 to 0.5. 
Silence intervals, randomly varying from 0.5 to 1.0 seconds, are inserted with a probability of 0.05.
To generate target speech signals, clean speech is convolved with the direct-path (anechoic) signal at the  reference microphone. To train the separation model, each mixture is divided into 5-second segments. Of these, 11\% contain a single speaker, and the remaining 89\% feature two. Prior to mixing the training utterances, we use our diarization model to extract an embedding vector for each speaker from the reverberant utterances of the speaker. This is done to ensure that embedding vectors accurately represent the corresponding speakers, as using mixture utterances may not produce adequate single-talker frames to extract quality speaker embeddings for training  purposes.

\subsection{Diarization Training and Evaluation Metrics}
For the EEND encoder, we employ eight stacked Transformer blocks, each equipped with 16 attention heads without  positional encodings. These encoder blocks generate $E=256$ dimensional frame-wise embeddings. The window size for STFT is set to 25 ms with a window shift of 10 ms, and the square root of the Hann window is used as the analysis window. A 512-point discrete Fourier transform is employed, resulting in the extraction of 257-dimensional complex spectra. For spectral features, we extract a 23-dimensional log-Mel filterbank, which is subsequently concatenated with those of the seven preceding and seven succeeding frames. As for spatial features, we process IPD features through seven convolutional blocks, having channel configurations of (8,8,16,16,32,32,32). These blocks use kernel sizes of (3, 5) and strides of (1, 2) along the time and frequency axes, respectively, except for the initial block which employs a kernel size of (15,1). These log-Mel features are subsequently concatenated with the spatial features. All input features are normalized to have zero mean and unit variance.

We use the Adam optimizer for training the diarization model with the learning rate of 0.001, which is coupled with the Noam scheduler~\cite{vaswani2017attention} including 125K warm-up steps.  We set the number of speakers to 8 for both training and testing. During diarization training, we use segments of 4-minute duration to ensure that all speakers are active within each segment. Sessions shorter than 4 minutes are zero padded at the end. To process long audio segments, we reduce  the number of frames in each batch via subsampling with a factor of 5, resulting in a 50-ms frame shift for training. For speech activity decisions, we use a threshold of  $\tau=0.5$. To avoid generating exceedingly short segments, a 31-frame median filter is applied. 

Diarization performance is evaluated using the NIST diarization error rate (DER)~\cite{sadjadi2021nist}, which calculates the combined durations of missed speech, false alarm, and speaker confusion errors, divided by the total duration of speech. We use a 0-second collar tolerance at utterance boundaries.

\subsection{Separation Training and Evaluation Metrics}
Our separation models employ the following DNN architectures:
\begin{itemize}
  \item  A TF-GridNet featuring 4 blocks, a 192-unit BLSTM, a kernel size of 4, a stride of 1, and $D=48$ channels.
  \item A large TF-GridNet with 6 blocks, a 256-unit BLSTM, and $D=64$ channels.
  \item A SpatialNet with 12 blocks, $D=192$ channels, narrow-band hidden dimensions of 384, and cross-band hidden dimensions of 16.
\end{itemize}

The STFT parameters for the TF-GridNet models are set to a window length of 32 ms and shift of 10 ms. For the SpatialNet model, the window length and shift are set to 32 ms and 16 ms, respectively. Both setups extract 257-dimensional complex spectra for the 16 kHz sampling rate.
We modify the TF-GridNet and SpatialNet models to incorporate speaker embedding sequences with a mixture signal. 
Specifically, we divide the input encoder layer, a two-dimensional convolution for TF-GridNet and a one-dimensional convolution for SpatialNet, each with $D$ channels, into two separate encoders. The two encoders, with $D-2$ and $2$ channels, process the mixture signal and the embedding sequences, respectively. The outputs of these two encoders are then stacked and fed into the subsequent blocks of each network.

Before training, the sample variance of each mixture segment is normalized to 1.0, and the corresponding scaling factor is applied to the clean target sources. We employ the Adam optimizer, with the $\ell_2$  norm of gradients capped at 1.0. The learning rate  is initialized to 0.001, and halved if no improvement in validation loss is observed over three epochs. All models use mixed precision to expedite training.

For the multi-speaker ASR evaluation, two pretrained ASR models from ESPnet are employed~\cite{watanabe2018espnet}. The first ASR model is an end-to-end Transformer-based system~\cite{e2eASR, rajSLTIntegration}. This model is equipped with 12 self-attention blocks in the encoder and 6 in the decoder, and trained on the 960h Librispeech corpus.
 The second ASR model\footnote{Available online at: \url{https://huggingface.co/espnet/simpleoier_librispeech_asr_train_asr_conformer7_wavlm_large_raw_en_bpe5000_sp}} is a conformer-based system, leveraging self-supervised learning (SSL) features derived from WavLM~\cite{chen2022wavlm}. This ASR model achieves a WER of 1.9\% on the clean test set of LibriSpeech. We refer to the first ASR model as E2E and the second as E2E-SSL.

 We report multi-speaker ASR performance using concatenated minimum-permutation word error rate (cpWER)~\cite{watanabe20b_chime}. This metric evaluates speaker-attributed ASR and is computed by sequentially joining all the utterances of each separated and target speaker. Following this joining, all speaker pairs are scored. The permutation yielding the lowest WER is then selected.

 We also conduct the continuous-input evaluation of LibriCSS~\cite{chen2020continuous}. For this evaluation, each 10-minute mini-session recording, is pre-segmented into segments spanning 60 to 120 seconds. Each of these segments includes 8 to 10 utterances. The objective here is to accurately recognize all the utterances within a segment in a speaker-agnostic manner. While the ASR backend evaluates both streams individually, it combines the decoding results to determine the final WER. For this evaluation, we employ the default ASR backend from the LibriCSS dataset for consistent comparisons with other algorithms.

\begin{table*}[t]
  \centering

  \caption{DER results (in \%) of comparison diarization systems on LibriCSS.}
  \renewcommand{\arraystretch}{1.25}
  \begin{adjustbox}{width=0.9\textwidth}
  \begin{tabular}{cl ccccccc}
      \toprule
      \multirow{2}{*}{Separation Method} & \multirow{2}{*}{Diarization Method} & \multicolumn{6}{c}{Overlap Ratio} & \multirow{2}{*}{All} \\
      \cmidrule(r){3-8}
      & &  0S & 0L & 10\% & 20\% & 30\% & 40\% \\

      \midrule
      -  & X-vector + SC~\cite{rajSLTIntegration} & 9.29 & 10.25 & 14.04 & 18.76 & 23.82 & 27.43 & 18.19 \\
      Mask-based MVDR~\cite{chen2020continuous} & X-vector + SC & 11.49 & 13.42 & 11.63 & 14.22 & 16.98 & 16.18 & 14.18 \\        
      MIMO-BF-MISO~\cite{taherianICASSP24} & X-vector + SC & 9.33 & 10.4 & 8.97 & 9.52 & 11.66 & 9.54 & 9.9 \\

      \midrule
      MISO-BF-MISO~\cite{wang2021multi} & DOA-based~\cite{ZQLocDiar22} & 11.95	&10.69	&11.25	&12.22	&13.04	&14.31	&12.36 \\
      -  & RPN~\cite{rajSLTIntegration} & 4.5 & 9.1 & 8.3 & 6.7 & 11.6 & 14.2 & 9.5 \\
      -  & TS-VAD~\cite{rajSLTIntegration} & 6.0 & \textbf{4.6} & 6.6 & 7.3 & 10.3 & 9.5 & 7.6 \\
      -  & TS-SEP~\cite{boeddeker2023ts} & - & - & - & - & - & - & 6.49 \\

      \midrule
      - & SC-EEND (PIT)                                    &23.2	&23.48	&29.55	&27.1	&34.94	&39.22	&30.36 \\
      - & MC-EEND (PIT)                                     &7.56	&6.57 & 5.09  &	7.18  &	8.44  &	12.17 &	8.05 \\
      - & MC-EEND (LBT)                                     &\textbf{4.94}	&6.12	&\textbf{3.36}	&\textbf{4.09}	&\textbf{4.88}	&\textbf{5.05} &	\textbf{4.68} \\

      \bottomrule
  \end{tabular}
  \end{adjustbox}
  \label{tab:diarization_results}
\end{table*}

\section{Evaluation Results and Comparisons}\label{sec_results}

\subsection{Diarization Results}

Table~\ref{tab:diarization_results} presents the DER results for the proposed MC-EEND models and comparison baselines on LibriCSS. To provide a comprehensive perspective on these results, we compare our MC-EEND models with a number of other diarization methods, all of which have been evaluated on the same LibriCSS corpus. The x-vector+SC diarization method~\cite{rajSLTIntegration} achieves 18.19\% DER. This diarization method can be combined with CSS-based separation for DER reduction. Here, clustering is performed for both separated streams concurrently, as speaker segments can be assigned to either separated stream.  Raj et al.~\cite{rajSLTIntegration} reported that using mask-based minimum variance distortionless response (MVDR) beamforming~\cite{chen2020continuous} prior to diarization improves DER results. 

We have conducted experiments using a more powerful CSS-based separation model. Specifically, we employ the MIMO-BF-MISO model~\cite{taherianICASSP24} for separation. This model is based on MIMO complex spectral mapping through a TF-GridNet architecture. Additionally, it includes a beamformer and an enhancement model for post-filtering. As part of  post-processing, the separation model performs speaker localization to reduce speaker splitting errors. Coupling this separation model and x-vector+SC cuts DER by half, resulting in a 8.29\% absolute DER reduction.

The next baseline employs a diarization model that relies on direction of arrival (DOA) estimation~\cite{ZQLocDiar22}. This model utilizes a CSS-based MISO-BF-MISO system~\cite{wang2021multi} to perform speaker separation, estimate the DOA for each separated speaker, and then group the separation results across segments according to the DOA estimates.
Region proposal network (RPN) is supervised method that integrates both segmentation and embedding extraction steps into one neural network, optimizing them jointly~\cite{RPN_diar}. After obtaining the embeddings, they are clustered using K-means clustering based on the oracle number of speakers.

Another notable diarization baseline is target-speaker voice activity detection (TS-VAD), a two-stage method~\cite{TSVAD_medennikov20_interspeech}. Initially, diarization estimates are obtained using a clustering-based method. Subsequently, a DNN is employed to refine these initial diarization estimates. Specifically, the DNN model takes acoustic features along with representative embeddings for each speaker as inputs and generates frame-level activities for each speaker. To address the permutation ambiguity problem, TS-VAD arranges the DNN outputs based on the order of input embeddings, and assumes the knowledge of the total number of speakers in a meeting. TS-VAD achieves strong diarization results in the CHiME-6 challenge~\cite{TSVAD_medennikov20_interspeech}.

An extension to TS-VAD is target-speaker separation (TS-SEP), which combines speaker diarization and separation into a unified process~\cite{boeddeker2023ts}. More specifically, TS-SEP extends the final output layer of TS-VAD to generate time-frequency masks for individual speakers. To achieve robust diarization and separation, TS-SEP incorporates several additional techniques. First, it employs weighted prediction error (WPE) for speech dereverberation~\cite{WPE} before mask estimation. Second, TS-SEP applies mask-based MVDR beamforming, and the resulting masks are further refined through guided source separation (GSS)~\cite{boeddecker18_chime}. This baseline achieves a 6.49\% DER.

\begin{figure}[!t]
  \centering
  \includegraphics[width=0.49\textwidth]{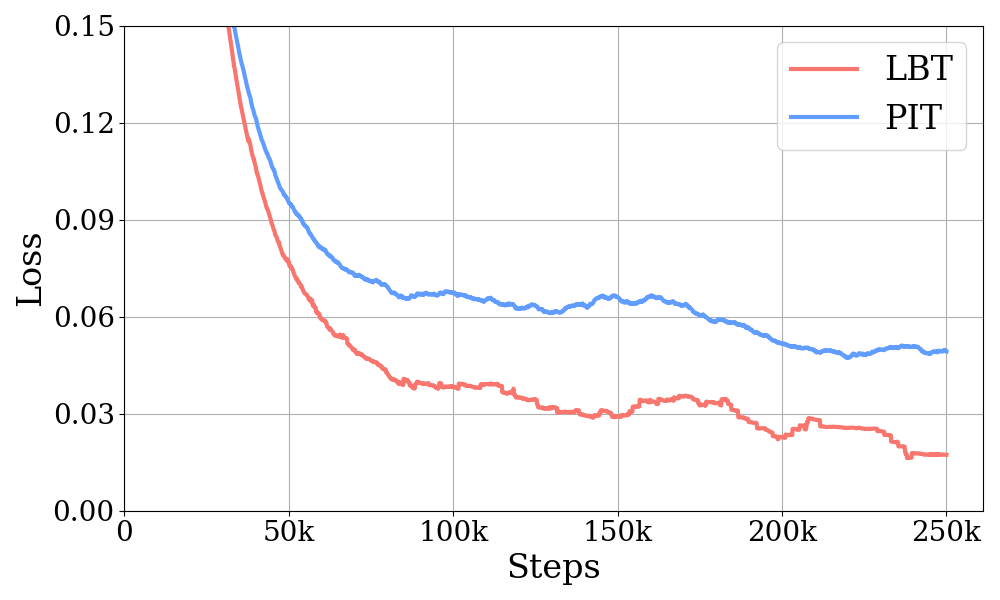}
  \caption{Comparison of diarization loss curves ($L_{\text{EEND}}$) for training with LBT and PIT criteria.}
  \label{fig_LBT_PIT_diar_loss}
  \end{figure}

Compared to all the aforementioned baselines, our proposed MC-EEND trained with LBT achieves a 4.68\% DER, surpassing all other diarization methods. It is worth stressing that MC-EEND attains these results without employing any speech separation  techniques. Furthermore, MC-EEND is a single-stage system, making it easier to train than multi-stage diarization methods that depend on other modules. The results also demonstrate that MC-EEND generalizes well to real conversational recordings, despite using only simulated RIRs for training.

For reference, we have also trained an MC-EEND diarization model using the PIT criterion. To accommodate training for 8 speakers, we employ a PIT criterion that leverages the Hungarian algorithm~\cite{dovrat2021many} with an $O(C^3)$ computational complexity. From Table~\ref{tab:diarization_results} we observe that the MC-EEND model with LBT  significantly outperforms that with PIT. Fig.~\ref{fig_LBT_PIT_diar_loss} displays the diarization loss curves for LBT and PIT, and it is evident from the curves that the diarization loss for LBT is considerably lower. This suggests that the PIT criterion may be suboptimal when dealing with a larger number of speakers.

To investigate this further, we have trained a single-channel EEND (SC-EEND), which is based on the original EEND-EDA~\cite{eendEDA}, using the PIT criterion. Table~\ref{tab:diarization_results} indicates that SC-EEND with the PIT criterion yields poor separation performance. This is consistent with the findings from other studies~\cite{eendEDA, FramewiseOverlapRobustDiar23} reporting the poor performance of PIT-based EEND models for many speakers.

\subsection{Diarization Tuning for SSND}

We extract speaker embeddings for SSND from our MC-EEND model trained with LBT. Prior to the extraction of embeddings, we fine-tune our MC-EEND model to achieve optimal cpWER results.
Of the three diarization errors — missed speech, confusion, and false alarm — the first two are particularly detrimental for ASR. Missed speech errors lead to deletion errors, while confusion errors result in a deletion error for one speaker and an insertion error for the other speaker. In contrast, false alarm errors do not contribute to cpWER because ASR systems do not generate recognition hypotheses for silent frames.
Additionally, we have empirically observed that overly strict segment boundaries contribute to deletion errors at the beginning and ending of an utterance. Therefore, tolerating false alarm errors can be a strategic choice to improve ASR performance.

\begin{table}[t]
  \centering
  \renewcommand{\arraystretch}{1.25}
  \caption{DER (in \%) for different frame shifts and diarization thresholds. MI, FA and CF refer to missed speech, false alarm, and confusion errors, respectively.
  }
  \begin{adjustbox}{width=0.48\textwidth}
  \begin{tabular}{cccccc}
  \toprule
  Frame Shift & Threshold & DER & MI & FA & CF \\ 
  \midrule
  30 ms & $\tau=0.5$ & \textbf{4.68} & 1.86 & \textbf{2.16} & 0.66 \\ 
  40 ms & $\tau=0.5$ & 4.97 & 1.51 & 2.87 & 0.59 \\ 
  50 ms & $\tau=0.5$ & 5.72 & 1.50 & 3.66 & 0.56 \\ 
  \midrule
  30 ms & $\tau=0.3$ & 7.98 & 0.84 & 6.48 & 0.66 \\ 
  40 ms & $\tau=0.3$ & 7.41 & 0.69 & 6.28 & 0.44 \\
  50 ms & $\tau=0.3$ & 7.62 & \textbf{0.53} & 6.98 & \textbf{0.11} \\
  \bottomrule
  \end{tabular}
  \end{adjustbox}
  \label{tab:diarization_tuning}
\end{table}

To enhance SSND performance, one can tune the diarization threshold ($\tau$) as well as frame shift.
Table~\ref{tab:diarization_tuning} provides diarization errors using different frame shifts and threshold values. From the table the best DER is obtained  at the smallest shift (30 ms) with $\tau=0.5$. With this setting,  missed speech and false alarm errors appear balanced. When we lower the threshold to 0.3, there is a marked reduction in missed speech errors, but at the expense of increased false alarm errors.

    \begin{figure}[t]
      \centering
      \includegraphics[width=0.45\textwidth]{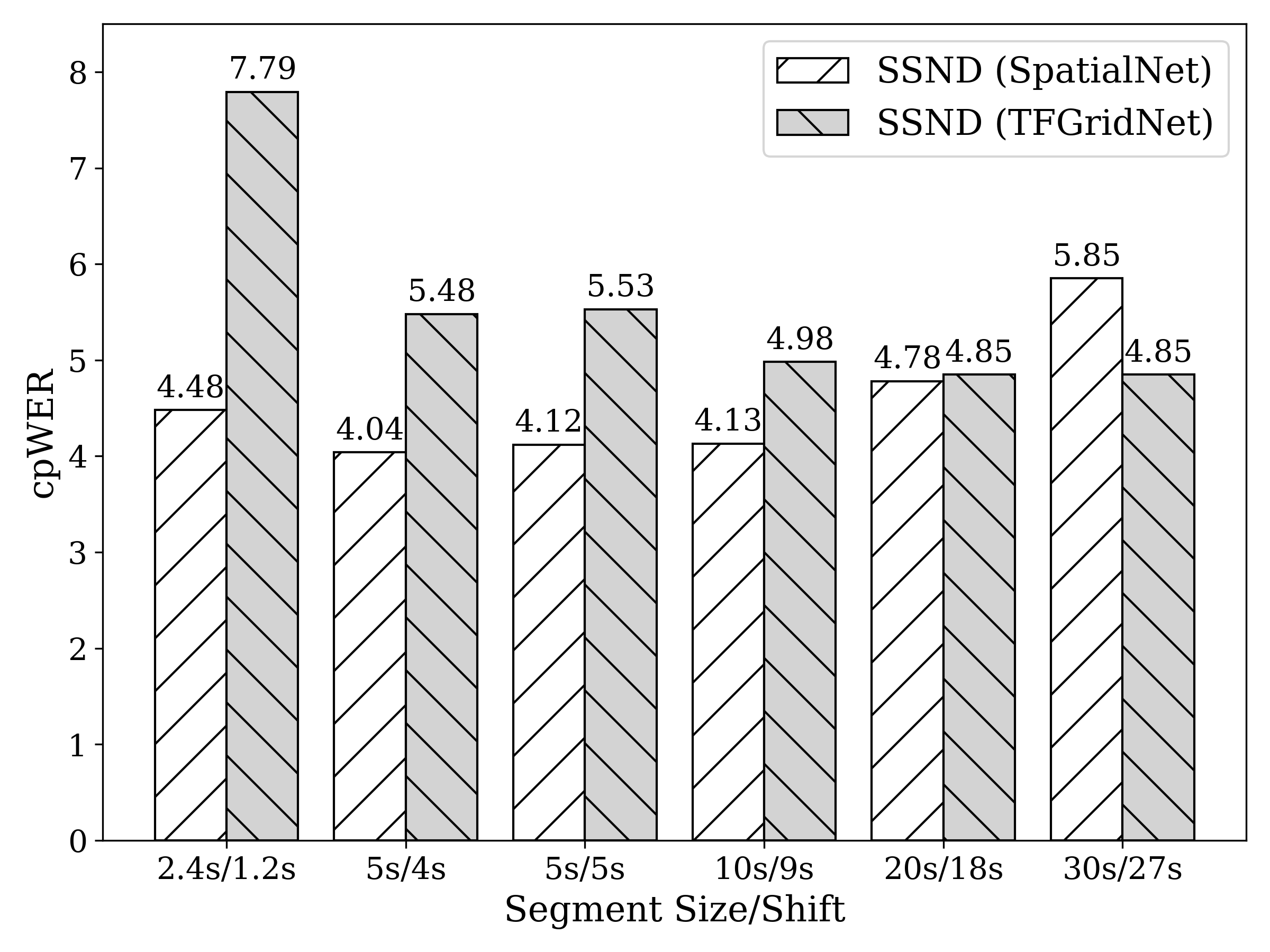}
      \caption{Effect of segment size and segment shift on cpWER in SSND.}
      \label{fig_SegAnalysis}
      \end{figure}

In terms of combined missed speech and confusion errors, the best setting is a 50 ms frame shift and $\tau=0.3$. Consequently, we adopt this MC-EEND setting for subsequent separation and ASR experiments.

  \begin{table*}[t]

    \centering
    \renewcommand{\arraystretch}{1.25} 
    \caption{cpWER results (in \%) for different separation and diarization methods.}
    \begin{adjustbox}{width=0.9\textwidth}
    \begin{tabular}{l c c cccccccc}
    \toprule
  
     \multirow{2}{*}{Separation Method} & \multirow{2}{*}{Diarization Method} & \multirow{2}{*}{ASR} & \multicolumn{6}{c}{Overlap Ratio} & \multirow{2}{*}{Avg} \\
     \cmidrule(r){4-9}
     & & & 0S & 0L & 10\% & 20\% & 30\% & 40\% \\
     \midrule
    Unprocessed & Oracle & E2E & 5.23 & 5.19 & 11.43 & 19.31 & 28.46 & 38.32 & 19.72 \\
    MIMO-BF-MISO~\cite{taherianICASSP24} & Oracle & E2E & 3.45	  &3.87	  &3.15	  &4.46	  &5.35 &5.76	&4.44\\
    SSND (TF-GridNet Large) & Oracle & E2E &  4.51& 4.01 & 3.74  & 4.28 & 5.13 & 5.39 & 4.58 \\

    SSND (SpatialNet) & Oracle & E2E &  4.04& 	3.97& 	3.37& 	3.54& 	4.51& 	4.66& 	4.04 \\
    SSND (TF-GridNet Large) & Oracle & E2E-SSL & 2.51 & 2.34 &  2.46 & 2.55 & 2.91 & 2.93 & 2.64 \\
    SSND (SpatialNet)  & Oracle & E2E-SSL &   2.28&	2.38&	 2.36&	2.27&	2.68&	2.47&	2.42 \\
    \midrule
    Unprocessed & X-vector + SC &E2E & 13.95 & 12.2 & 20.12 & 29.64 & 35.06 & 41.81 & 27.01 \\
    Mask-based MVDR~\cite{chen2020continuous} & X-vector + SC &E2E & 8.78 & 13.07 & 10.51 & 15.37 & 17.54 & 17.63 & 14.13 \\
    MIMO-BF-MISO~\cite{taherianICASSP24} & X-vector + SC &E2E & 7.05&	8.05&	7.31&	8.48&	10.81&	10.03&	8.76 \\
    SSND (TF-GridNet Large) & MC-EEND $(\tau=0.5)$ &E2E &   5.11	& 4.88	&5.55	& 6.89	&  7.98	& 9.25	&  6.84	\\
    SSND (TF-GridNet Large) & MC-EEND $(\tau=0.3)$ &E2E &  5.97	& 3.74	&4.69 &	5.32 &	 6.18&	7.28 	&5.69\\

    SSND (SpatialNet) & MC-EEND $(\tau=0.3)$ &E2E & 5.56& 	 3.52	& 3.98	& 4.76& 	5.58	& 6.55	& 5.13 \\
    SSND (TF-GridNet Large) & MC-EEND $(\tau=0.3)$ & E2E-SSL & 3.77 & 2.02  & 2.56 & 3.29 & 3.78 &  4.11 & 3.36 \\
    SSND (SpatialNet) & MC-EEND $(\tau=0.3)$ &E2E-SSL & 3.67	&1.97&	2.46&	3.06&	3.52&4.07&	3.22 \\
  
    \bottomrule
    \end{tabular}
    \end{adjustbox}
    \label{tab:bigTable}
    \end{table*}
  
\subsection{Segment Size and Shift Analysis for SSND}
  
In this section, we investigate the influence of segment size and shift on the performance of SSND. This analysis utilizes the TF-GridNet and SpatialNet architectures, and an E2E ASR method with oracle utterance boundaries. The cpWER results for various segment sizes and shifts  processed through SSND are shown in Fig.~\ref{fig_SegAnalysis}.

The default CSS segment size and shift (2.4/1.2 seconds) offer limited contextual information and lead to subpar cpWER performance, particularly with TF-GridNet. It is important to note that our approach uses no stitching, meaning these cpWERs are not due to potential stitching errors. For SSND using the TF-GridNet model, increasing the segment size up to 30 seconds (i.e. 12 times the default CSS segment size) results in improved performance. This finding is significant as it suggests the model ability to process longer segments without compromising performance. Interestingly, in scenarios with no overlap between adjacent segments, e.g., segment size/shift of 5/5 seconds, cpWER remains good. This observation implies that one can reduce the computational cost for SSND with little performance degradation. In the case of SSND with SpatialNet, increasing the segment size improves ASR performance up to 10 seconds. For longer segments (20 seconds and 30 seconds), performance starts to decline. This trend may be attributed to the lack of a full-band self-attention module and recurrent layers in SpatialNet to capture long-range contextual information.
Based on the results in Fig.~\ref{fig_SegAnalysis}, we adopt a segment size/shift of 30/27 seconds for TF-GridNet models and 5/4 seconds for SpatialNet in the subsequent experiments.

\subsection{Speaker-attributed ASR Results}


In this section, we present the speaker-attributed ASR results on the LibriCSS dataset. Table~\ref{tab:bigTable} displays the cpWER results using various separation and diarization methods. For comparisons, we establish a strong baseline using a CSS method based on MIMO-BF-MISO built on TF-GridNet~\cite{taherianICASSP24}.
To obtain cpWER results with oracle utterance boundaries for the MIMO-BF-MISO system, the cross-correlation function between the separated audio streams and the reference utterances is used to determine channel assignments accurately.

A notable observation is that the MIMO-BF-MISO method reduces cpWER by a large margin --- from 19.72\% to 4.44\% --- over the unprocessed mixtures when coupled with E2E ASR using oracle utterance boundaries. Furthermore, SSND with TF-GridNet Large produces ASR results on par with MIMO-BF-MISO, while SSND with SpatialNet surpasses the performance of both models.
Using E2E-SSL ASR, we achieve excellent recognition results for the TF-GridNet and SpatialNet SSND models, with 2.64\% and 2.42\% cpWER respectively. These results are close to the clean transcription level of 1.9\%. Interestingly, we  observe that the difference in performance between TF-GridNet and SpatialNet becomes marginal when E2E-SSL is used for ASR.

Using estimated speaker boundaries provided by  diarization, we notice a significant increase in cpWER for the MIMO-BF-MISO system coupled with clustering-based diarization, to an 8.76\% cpWER. However, this performance remains substantially better than the mask-based MVDR result at 14.13\% cpWER.

For SSND using the TF-GridNet Large model, there is a small increase in cpWER to 5.69\%  over oracle speaker boundaries. The impact of diarization tuning becomes evident when an  MC-EEND diarization model is employed with $\tau=0.5$, degrading cpWER to 6.84\%. SSND with SpatialNet surpasses the performance of TF-GridNet Large, reaching a 5.13\% cpWER. Moreover, using E2E-SSL for ASR yields a marked improvement in cpWER for both TF-GridNet Large and SpatialNet models, achieving cpWERs of 3.36\% and 3.22\% respectively.

\begin{table}[t]
  \centering
  \caption{Performance Comparisons of Speaker-attributed ASR Systems on LibriCSS Dataset.}
  \renewcommand{\arraystretch}{1.25} 
  \begin{adjustbox}{width=0.49\textwidth}
  \begin{tabular}{ccccc}
      \toprule
      Ref. & Separation &Diarization & ASR & cpWER \\
      \midrule
      \cite{ZQLocDiar22} & CSS & DOA-based & TDNN-F~\cite{rajSLTIntegration} & 12.98 \\
      \cite{rajSLTIntegration} & CSS & X-vector + SC  & E2E & 12.7 \\
      \cite{Transcribe2Diariaze} & – & – & SA-ASR & 11.6 \\
      \cite{TSVAD_beam} & Speakerbeam & TS-VAD & E2E & 18.8 \\
      \cite{TSVAD_beam} & GSS & TS-VAD & E2E & 11.2 \\
      \midrule
      \cite{boeddeker2023ts} &\multicolumn{2}{c}{TS-SEP} & E2E & 6.42 \\
      \cite{boeddeker2023ts} &\multicolumn{2}{c}{TS-SEP} & E2E-SSL  & 5.36 \\
      \midrule
      Ours &  \multicolumn{2}{c}{SSND (SpatialNet)} & E2E & 5.13 \\
      Ours & \multicolumn{2}{c}{SSND (SpatialNet)} & E2E-SSL & \textbf{3.22} \\
      \bottomrule
  \end{tabular}
    \end{adjustbox}
  \label{tab:literature_comparison}
\end{table}


We compare the speaker-attributed performance of the proposed SSND model with other representative algorithms in Table~\ref{tab:literature_comparison}. The CSS-based MISO-BF-MISO system coupled with DOA-based diarization and a hybrid ASR model~\cite{rajSLTIntegration} reports a 12.98\% cpWER~\cite{ZQLocDiar22}. Another system outlined in~\cite{rajSLTIntegration} employs mask-based MVDR, x-vector+SC diarization, and E2E ASR, and achieves a 12.7\% cpWER~\cite{rajSLTIntegration}. A single-channel E2E speaker-attributed ASR system~\cite{Transcribe2Diariaze} derives diarization estimates from the internal state of the recognizer, and achieves a 11.6\% cpWER.
The system in~\cite{TSVAD_beam} uses TS-VAD for diarization and reports the cpWER scores of 18.8\% and 11.2\% for single- and multi-channel setups utilizing a speakerbeam and GSS, respectively. 
Finally, the TS-SEP system~\cite{boeddeker2023ts} reaches the cpWER values of 6.42\% and 5.36\% with E2E and E2E-SSL ASR, respectively. The cpWER results in Table~\ref{tab:literature_comparison} demonstrate that our proposed SSND model surpasses all previous results, achieving the remarkable cpWER scores of 5.13\% and 3.22\% for E2E and E2E-SSL, respectively. Our results establish a new state-of-the-art benchmark for speaker-attributed ASR on the LibriCSS dataset.

\subsection{Speaker-agnostic ASR Results}

This section  assesses our SSND model using the continuous-input evaluation of the LibriCSS dataset. Additionally, we compare with other works using the default ASR backend. It is important to note that these results do not incorporate diarization since the segment boundaries (with several utterance from different speakers) are provided for this evaluation, so the corresponding ASR is referred to as speaker-agnostic. The system in~\cite{Chen2021conformer} estimates real-valued time-frequency masks using a conformer architecture.

Table~\ref{tab:spk_agnostic_ASR} presents the continuous-input evaluation results of the proposed SSND and comparison methods. The results show that our SSND based on TF-GridNet significantly outperforms the corresponding CSS-based system. This can be attributed to the utilization of a larger context—30 seconds as opposed to the 2.4 seconds used in CSS.
Using SpatialNet, our SSND results show a further improvement over TF-GridNet, achieving 7.33\% WER on average and surpassing the performance level of the best-performing CSS system using MIMO-BF-MISO~\cite{taherianICASSP24}. It should be noted that, unlike MIMO-BF-MISO, the SSND system does not perform additional processing such as beamforming and localization. On the other hand, SSND makes use of speaker embedding sequences as additional inputs.

\begin{table}[t]
  \centering
  \caption{WER results (in \%) of comparison systems for continuous-input evaluation on LibriCSS using default ASR backend.}
  \label{table:your_label}
  \renewcommand{\arraystretch}{1.25}
  \begin{adjustbox}{width=0.49\textwidth}
  \begin{tabular}{lccccccc}
      \toprule
      \multirow{2}{*}{} & \multicolumn{6}{c}{Overlap Ratio} & \multirow{2}{*}{Avg}\\
      \cmidrule(r){2-7}
      & 0S & 0L & 10\% & 20\% & 30\% & 40\% & \\
      \midrule
      Unprocessed & 15.4 & 11.5 & 21.7 & 27 & 34.3 & 40.5 & 25.06 \\
      \midrule
      \multicolumn{8}{l}{\textit{\textbf{CSS}}} \\
      \hspace{3mm}MVDR~\cite{chen2020continuous} & 11.9 & 9.7 & 13.4 & 15.1 & 19.7 & 22.0 & 15.29\\
      \hspace{3mm}Conformer~\cite{Chen2021conformer} & 11.0 & 8.7 & 12.6 & 13.5 & 17.6 & 19.6 & 13.83 \\
      \hspace{3mm}MISO-BF-MISO~\cite{wang2021multi} & 7.7 & 7.5 & 7.4 & 8.4 & 9.7 & 11.3 & 8.66 \\
      \hspace{3mm}TF-GridNet & 9.0 & 10.8 & 10 & 10.4 & 12.0 & 12.9 & 10.85 \\
      \hspace{3mm}MIMO-BF-MISO~\cite{taherianICASSP24} & \textbf{6.8} & 6.8 & \textbf{6.7} & 6.9 & 8.4 & 9.0 & 7.43 \\
      \midrule
      \multicolumn{8}{l}{\textit{\textbf{SSND}}} \\
      \hspace{3mm}TF-GridNet & 7.9 & 7.4 & 7.8 & 7.8 & 9.8 & 10.3 & 8.50 \\

      \hspace{3mm}SpatialNet & 7.2 & \textbf{6.5} & \textbf{6.7}  & \textbf{6.6} & \textbf{8.3} & \textbf{8.6} & \textbf{7.33} \\

      \bottomrule
  \end{tabular}
  \end{adjustbox}
  \label{tab:spk_agnostic_ASR}
\end{table}

\section{Concluding Remarks}\label{section_conclude}

In this paper, we have proposed a new multi-channel diarization model, MC-EEND, which produces state-of-the-art diarization performance on the LibriCSS dataset. We find that PIT falls short when diarizing many speakers. With multi-channel recordings, we demonstrate that the LBT criterion effectively resolves the permutation ambiguity problem  in talker-independent diarization.

Furthermore, we have introduced the SSND framework, a novel approach that seamlessly integrates speaker diarization with speaker separation, making it well-suited for speaker-attributed ASR. Our SSND framework achieves state-of-the-art performance for speaker-attributed ASR, as well as speaker-agnostic ASR (standard CSS),  on the LibriCSS dataset.  Unlike CSS, the SSND framework is capable of processing long  segments regardless the number of participating speakers. SSND avoids stitching needed in CSS and ensures that consecutive segments are sequentially organized.

Future research will extend MC-EEND to causal and real-time implementation and moving speakers, and connect to speaker localization and tracking. Additional research is also needed to deal with many speakers in a single-channel setup.

\section*{Acknowledgments}
The authors would like to thank Dr. Zhong-Qiu Wang, Dr. Ashutosh Pandey, Dr. Daniel Wong and Dr. Buye Xu for helpful discussions. This research was supported in part by an National Science Foundation grant (ECCS-2125074), a research contract from Meta Reality Labs, the Ohio Supercomputer Center, and the Pittsburgh Supercomputer Center (NSF ACI-1928147).

\bibliographystyle{IEEEtran}
{\small\bibliography{mybib}}

\begin{thebibliography}{10}
\providecommand{\url}[1]{#1}
\csname url@samestyle\endcsname
\providecommand{\newblock}{\relax}
\providecommand{\bibinfo}[2]{#2}
\providecommand{\BIBentrySTDinterwordspacing}{\spaceskip=0pt\relax}
\providecommand{\BIBentryALTinterwordstretchfactor}{4}
\providecommand{\BIBentryALTinterwordspacing}{\spaceskip=\fontdimen2\font plus
\BIBentryALTinterwordstretchfactor\fontdimen3\font minus
  \fontdimen4\font\relax}
\providecommand{\BIBforeignlanguage}[2]{{%
\expandafter\ifx\csname l@#1\endcsname\relax
\typeout{** WARNING: IEEEtran.bst: No hyphenation pattern has been}%
\typeout{** loaded for the language `#1'. Using the pattern for}%
\typeout{** the default language instead.}%
\else
\language=\csname l@#1\endcsname
\fi
#2}}
\providecommand{\BIBdecl}{\relax}
\BIBdecl

\bibitem{wang2018supervised}
D.~L. Wang and J.~Chen, ``Supervised speech separation based on deep learning:
  An overview,'' \emph{IEEE/ACM Trans. Audio, Speech, Lang. Process.}, vol.~26,
  pp. 1702--1726, 2018.

\bibitem{chen2020continuous}
Z.~Chen, T.~Yoshioka, L.~Lu, T.~Zhou, Z.~Meng, Y.~Luo, J.~Wu, X.~Xiao, and
  J.~Li, ``Continuous speech separation: Dataset and analysis,'' in \emph{Proc.
  ICASSP}, 2020, pp. 7284--7288.

\bibitem{taherian2021time}
H.~Taherian and D.~L. Wang, ``Time-domain loss modulation based on overlap
  ratio for monaural conversational speaker separation,'' in \emph{Proc.
  ICASSP}, 2021, pp. 5744--5748.

\bibitem{DontShootButterfly}
S.~Chen, Y.~Wu, Z.~Chen, T.~Yoshioka, S.~Liu, J.~Li, and X.~Yu, ``Don’t shoot
  butterfly with rifles: Multi-channel continuous speech separation with early
  exit transformer,'' in \emph{Proc. ICASSP}, 2021, pp. 6139--6143.

\bibitem{Chenda}
C.~Li, Z.~Chen, and Y.~Qian, ``Dual-path modeling with memory embedding model
  for continuous speech separation,'' \emph{IEEE/ACM Trans. Audio, Speech,
  Lang. Process.}, vol.~30, pp. 1508--1520, 2022.

\bibitem{wang2021multi}
Z.-Q. Wang, P.~Wang, and D.~L. Wang, ``Multi-microphone complex spectral
  mapping for utterance-wise and continuous speech separation,'' \emph{IEEE/ACM
  Trans. Audio, Speech, Lang. Process.}, pp. 2001--2014, 2021.

\bibitem{taherian2023multi}
H.~Taherian and D.~L. Wang, ``Multi-resolution location-based training for
  multi-channel continuous speech separation,'' in \emph{Proc. ICASSP}, 2023,
  pp. 1--5.

\bibitem{taherian23_interspeech}
H.~Taherian, A.~Pandey, D.~Wong, B.~Xu, and D.~L. Wang, ``Multi-input
  multi-output complex spectral mapping for speaker separation,'' in
  \emph{Proc. Interspeech}, 2023, pp. 1070--1074.

\bibitem{kolbaek2017multitalker}
M.~Kolb{\ae}k, D.~Yu, Z.-H. Tan, and J.~Jensen, ``Multitalker speech separation
  with utterance-level permutation invariant training of deep recurrent neural
  networks,'' \emph{IEEE/ACM Trans. Audio, Speech, Lang. Process.}, vol.~25,
  pp. 1901--1913, 2017.

\bibitem{GraphPIT}
T.~v. Neumann, K.~Kinoshita, C.~Boeddeker, M.~Delcroix, and R.~Haeb-Umbach,
  ``Segment-less continuous speech separation of meetings: Training and
  evaluation criteria,'' \emph{IEEE/ACM Trans. Audio, Speech, Lang. Process.},
  vol.~31, pp. 576--589, 2023.

\bibitem{zhang22y_interspeech}
W.~Zhang, Z.~Chen, N.~Kanda, S.~Liu, J.~Li, S.~{Emre Eskimez}, T.~Yoshioka,
  X.~Xiao, Z.~Meng, Y.~Qian, and F.~Wei, ``Separating long-form speech with
  group-wise permutation invariant training,'' in \emph{Proc. Interspeech},
  2022, pp. 5383--5387.

\bibitem{eendEDA}
S.~Horiguchi, Y.~Fujita, S.~Watanabe, Y.~Xue, and P.~García, ``Encoder-decoder
  based attractors for end-to-end neural diarization,'' \emph{IEEE/ACM Trans.
  Audio Speech Lang. Process.}, vol.~30, pp. 1493--1507, 2022.

\bibitem{taherian2022LBTJorunal}
H.~Taherian, K.~Tan, and D.~L. Wang, ``Multi-channel talker-independent speaker
  separation through location-based training,'' \emph{IEEE/ACM Trans. Audio
  Speech Lang. Process.}, vol.~30, pp. 2791--2800, 2022.

\bibitem{diarReview2006}
S.~Tranter and D.~Reynolds, ``An overview of automatic speaker diarization
  systems,'' \emph{IEEE Trans. Audio, Speech, Lang. Process.}, vol.~14, pp.
  1557--1565, 2006.

\bibitem{park2022review}
T.~J. Park, N.~Kanda, D.~Dimitriadis, K.~J. Han, S.~Watanabe, and S.~Narayanan,
  ``A review of speaker diarization: Recent advances with deep learning,''
  \emph{Computer Speech \& Language}, vol.~72, p. 101317, 2022.

\bibitem{snyder2018x}
D.~Snyder, D.~Garcia-Romero, G.~Sell, D.~Povey, and S.~Khudanpur, ``X-vectors:
  Robust {DNN} embeddings for speaker recognition,'' in \emph{Proc. ICASSP},
  2018, pp. 5329--5333.

\bibitem{Fujita2019Interspeech}
Y.~Fujita, N.~Kanda, S.~Horiguchi, K.~Nagamatsu, and S.~Watanabe, ``End-to-end
  neural speaker diarization with permutation-free objectives,'' in \emph{Proc.
  Interspeech}, 2019, pp. 4300--4304.

\bibitem{KeSpectrospatial}
K.~Tan, Z.-Q. Wang, and D.~L. Wang, ``Neural spectrospatial filtering,''
  \emph{IEEE/ACM Trans. Audio, Speech, Lang. Process.}, vol.~30, pp. 605--621,
  2022.

\bibitem{hershey2016deep}
J.~R. Hershey, Z.~Chen, J.~Le~Roux, and S.~Watanabe, ``Deep clustering:
  Discriminative embeddings for segmentation and separation,'' in \emph{Proc.
  ICASSP}, 2016, pp. 31--35.

\bibitem{tachibana2021towards}
H.~Tachibana, ``Towards listening to 10 people simultaneously: An efficient
  permutation invariant training of audio source separation using sinkhorn’s
  algorithm,'' in \emph{Proc. ICASSP}, 2021, pp. 491--495.

\bibitem{dovrat2021many}
S.~Dovrat, E.~Nachmani, and L.~Wolf, ``Many-speakers single channel speech
  separation with optimal permutation training,'' in \emph{Proc. Interspeech},
  2021, pp. 3890--3894.

\bibitem{cetin06_interspeech}
Özgür Çetin and E.~Shriberg, ``{Analysis of overlaps in meetings by dialog
  factors, hot spots, speakers, and collection site: insights for automatic
  speech recognition},'' in \emph{Proc. ICSLP}, 2006, pp. 293--296.

\bibitem{wang2023tf}
Z.-Q. Wang, S.~Cornell, S.~Choi, Y.~Lee, B.-Y. Kim, and S.~Watanabe,
  ``{TF-GridNet}: Integrating full-and sub-band modeling for speech
  separation,'' \emph{IEEE/ACM Trans. Audio Speech Lang. Process.}, vol.~31,
  pp. 3221--3236, 2023.

\bibitem{quan2023spatialnet}
C.~Quan and X.~Li, ``{SpatialNet}: Extensively learning spatial information for
  multichannel joint speech separation, denoising and dereverberation,''
  \emph{arXiv:2307.16516}, 2023.

\bibitem{williamson2016complex}
D.~S. Williamson, Y.~Wang, and D.~L. Wang, ``Complex ratio masking for monaural
  speech separation,'' \emph{IEEE/ACM Trans. Audio, Speech, Lang. Process.},
  vol.~24, pp. 483--492, 2016.

\bibitem{9018157}
Z.-Q. Wang and D.~L. Wang, ``Deep learning based target cancellation for speech
  dereverberation,'' \emph{IEEE/ACM Trans. Audio, Speech, Lang. Process.},
  vol.~28, pp. 941--950, 2020.

\bibitem{OneModel}
H.~Taherian, S.~E. Eskimez, T.~Yoshioka, H.~Wang, Z.~Chen, and X.~Huang, ``One
  model to enhance them all: Array geometry agnostic multi-channel personalized
  speech enhancement,'' in \emph{Proc. ICASSP}, 2022, pp. 271--275.

\bibitem{zeghidour2021wavesplit}
N.~Zeghidour and D.~Grangier, ``Wavesplit: End-to-end speech separation by
  speaker clustering,'' \emph{IEEE/ACM Trans. Audio, Speech, Lang. Process.},
  vol.~29, pp. 2840--2849, 2021.

\bibitem{LibrispeechDataset}
V.~Panayotov, G.~Chen, D.~Povey, and S.~Khudanpur, ``Librispeech: An {ASR}
  corpus based on public domain audio books,'' in \emph{Proc. ICASSP}, 2015,
  pp. 5206--5210.

\bibitem{vaswani2017attention}
A.~Vaswani, N.~Shazeer, N.~Parmar, J.~Uszkoreit, L.~Jones, A.~N. Gomez,
  {\L}.~Kaiser, and I.~Polosukhin, ``Attention is all you need,'' \emph{Proc.
  NIPS}, vol.~30, 2017.

\bibitem{sadjadi2021nist}
O.~Sadjadi, C.~Greenberg, E.~Singer, L.~Mason, and D.~Reynolds, ``{NIST} 2021
  speaker recognition evaluation plan,'' Technical Report, 2021, available
  online at
  \url{https://tsapps.nist.gov/publication/get_pdf.cfm?pub_id=932697}.

\bibitem{watanabe2018espnet}
S.~Watanabe, T.~Hori, S.~Karita, T.~Hayashi, J.~Nishitoba, Y.~Unno, N.~Yalta,
  J.~Heymann, M.~Wiesner, N.~Chen, A.~Renduchintala, and T.~Ochiai, ``{ESPnet}:
  End-to-end speech processing toolkit,'' in \emph{Proceedings of Interspeech},
  2018, pp. 2207--2211.

\bibitem{e2eASR}
S.~Karita, N.~Chen, T.~Hayashi, T.~Hori, H.~Inaguma, Z.~Jiang, M.~Someki,
  N.~E.~Y. Soplin, R.~Yamamoto, X.~Wang, S.~Watanabe, T.~Yoshimura, and
  W.~Zhang, ``A comparative study on transformer vs {RNN} in speech
  applications,'' in \emph{Proc. ASRU}, 2019, pp. 449--456.

\bibitem{rajSLTIntegration}
D.~Raj, P.~Denisov, Z.~Chen, H.~Erdogan, Z.~Huang, M.~He, S.~Watanabe, J.~Du,
  T.~Yoshioka, Y.~Luo, N.~Kanda, J.~Li, S.~Wisdom, and J.~R. Hershey,
  ``Integration of speech separation, diarization, and recognition for
  multi-speaker meetings: System description, comparison, and analysis,'' in
  \emph{IEEE Spoken Language Technology Workshop}, 2021, pp. 897--904.

\bibitem{chen2022wavlm}
S.~Chen, C.~Wang, Z.~Chen, Y.~Wu, S.~Liu, Z.~Chen, J.~Li, N.~Kanda,
  T.~Yoshioka, X.~Xiao \emph{et~al.}, ``Wavlm: Large-scale self-supervised
  pre-training for full stack speech processing,'' \emph{IEEE Journal of
  Selected Topics in Signal Processing}, vol.~16, pp. 1505--1518, 2022.

\bibitem{watanabe20b_chime}
S.~Watanabe, M.~Mandel, J.~Barker, E.~Vincent, A.~Arora, X.~Chang,
  S.~Khudanpur, V.~Manohar, D.~Povey, D.~Raj, D.~Snyder, A.~S. Subramanian,
  J.~Trmal, B.~B. Yair, C.~Boeddeker, Z.~Ni, Y.~Fujita, S.~Horiguchi, N.~Kanda,
  T.~Yoshioka, and N.~Ryant, ``{CHiME}-6 challenge: Tackling multispeaker
  speech recognition for unsegmented recordings,'' in \emph{Proc. International
  Workshop on Speech Processing in Everyday Environments}, 2020, pp. 1--7.

\bibitem{taherianICASSP24}
H.~Taherian, A.~Pandey, D.~Wong, B.~Xu, and D.~L. Wang, ``Leveraging sound
  localization to improve continuous speaker separation,'' submitted to
  \textit{ICASSP} 2024, 2023.

\bibitem{ZQLocDiar22}
Z.-Q. Wang and D.~Wang, ``Localization based sequential grouping for continuous
  speech separation,'' in \emph{Proc. ICASSP}, 2022, pp. 281--285.

\bibitem{boeddeker2023ts}
C.~Boeddeker, A.~S. Subramanian, G.~Wichern, R.~Haeb-Umbach, and J.~L. Roux,
  ``{TS-SEP}: Joint diarization and separation conditioned on estimated speaker
  embeddings,'' \emph{arXiv:2303.03849}, 2023.

\bibitem{RPN_diar}
Z.~Huang, S.~Watanabe, Y.~Fujita, P.~García, Y.~Shao, D.~Povey, and
  S.~Khudanpur, ``Speaker diarization with region proposal network,'' in
  \emph{Proc. ICASSP}, 2020, pp. 6514--6518.

\bibitem{TSVAD_medennikov20_interspeech}
I.~Medennikov, M.~Korenevsky, T.~Prisyach, Y.~Khokhlov, M.~Korenevskaya,
  I.~Sorokin, T.~Timofeeva, A.~Mitrofanov, A.~Andrusenko, I.~Podluzhny,
  A.~Laptev, and A.~Romanenko, ``Target-speaker voice activity detection: A
  novel approach for multi-speaker diarization in a dinner party scenario,'' in
  \emph{Proc. Interspeech}, 2020, pp. 274--278.

\bibitem{WPE}
T.~Nakatani, T.~Yoshioka, K.~Kinoshita, M.~Miyoshi, and B.-H. Juang, ``Speech
  dereverberation based on variance-normalized delayed linear prediction,''
  \emph{IEEE Trans. Audio, Speech, Lang. Process.}, vol.~18, pp. 1717--1731,
  2010.

\bibitem{boeddecker18_chime}
C.~Boeddecker, J.~Heitkaemper, J.~Schmalenstroeer, L.~Drude, J.~Heymann, and
  R.~Haeb-Umbach, ``{Front-end processing for the CHiME-5 dinner party
  scenario},'' in \emph{Proc. International Workshop on Speech Processing in
  Everyday Environments}, 2018, pp. 35--40.

\bibitem{FramewiseOverlapRobustDiar23}
T.~Cord-Landwehr, C.~Boeddeker, C.~Zorilă, R.~Doddipatla, and R.~Haeb-Umbach,
  ``Frame-wise and overlap-robust speaker embeddings for meeting diarization,''
  in \emph{Proc. ICASSP}, 2023, pp. 1--5.

\bibitem{Transcribe2Diariaze}
N.~Kanda, X.~Xiao, Y.~Gaur, X.~Wang, Z.~Meng, Z.~Chen, and T.~Yoshioka,
  ``Transcribe-to-diarize: Neural speaker diarization for unlimited number of
  speakers using end-to-end speaker-attributed {ASR},'' in \emph{Proc. ICASSP},
  2022, pp. 8082--8086.

\bibitem{TSVAD_beam}
M.~Delcroix, K.~Zmolikova, T.~Ochiai, K.~Kinoshita, and T.~Nakatani, ``Speaker
  activity driven neural speech extraction,'' in \emph{Proc. ICASSP}, 2021, pp.
  6099--6103.

\bibitem{Chen2021conformer}
S.~Chen, Y.~Wu, Z.~Chen, J.~Wu, J.~Li, T.~Yoshioka, C.~Wang, S.~Liu, and
  M.~Zhou, ``Continuous speech separation with conformer,'' in \emph{Proc.
  ICASSP}, 2021, pp. 5749--5753.

\end{thebibliography}



\end{document}